# A Unifying Framework for Linearly Solvable Control

**Krishnamurthy Dvijotham**
Computer Science and Engg
University of Washington
Seattle, WA 98195

**Emanuel Todorov**
Computer Science and Engg & Applied Mathematics
University of Washington
Seattle, WA 98195

## Abstract

Recent work has led to the development of an elegant theory of Linearly Solvable Markov Decision Processes (**LMDPs**) and related Path-Integral Control Problems. Traditionally, **LMDPs** have been formulated using stochastic policies and a control cost based on the KL divergence. In this paper, we extend this framework to a more general class of divergences: the Rényi divergences. These are a more general class of divergences parameterized by a continuous parameter $\alpha$ that include the KL divergence as a special case. The resulting control problems can be interpreted as solving a risk-sensitive version of the **LMDP** problem. For $\alpha > 0$, we get risk-averse behavior (the degree of risk-aversion increases with $\alpha$) and for $\alpha < 0$, we get risk-seeking behavior. We recover **LMDPs** in the limit as $\alpha \to 0$. This work generalizes the recently developed risk-sensitive path-integral control formalism which can be seen as the continuous-time limit of results obtained in this paper. To the best of our knowledge, this is a general theory of linearly solvable control and includes all previous work as a special case. We also present an alternative interpretation of these results as solving a 2-player (cooperative or competitive) Markov Game. From the linearity follow a number of nice properties including compositionality of control laws and a path-integral representation of the value function. We demonstrate the usefulness of the framework on control problems with noise where different values of $\alpha$ lead to qualitatively different control behaviors.

## 1 INTRODUCTION

Optimal Control is a conceptually appealing framework for building solutions to complex control problems. However, computational intractability has severely limited its application to practical problems with nonlinear dynamics and high dimensional continuous state/control spaces. In recent years, researchers have developed restricted, yet fairly general classes of control problems that are more tractable: examples include *Linearly Solvable MDPs (***LMDPs***)* (Todorov, 2009) and related path-integral control problems (**PIC**) (Kappen, 2005), for which the Bellman equation characterizing the optimal value function can be made linear. This has several other interesting consequences: the ability to build solutions to new control problems by combining the solutions to simpler control problems (Todorov, 2009), using probabilistic inference techniques for control (Mensink *et al.*, 2010) etc. Already, this work has had encouraging success in domains like character control for animation (Da Silva *et al.*, 2009) and robotic control (Theodorou *et al.*, 2010). In this paper, we present a more general class of linearly solvable control problems.

Traditional MDPs are formulated as minimizing expected accumulated costs (over a finite or infinite time horizon). However, in many applications, one cares about higher order moments of the accumulated cost (like its variance) that depend on the amount of noise in the system. This is particularly relevant for noisy underactuated systems near unstable equilibria, since it can be hard to recover from even small amounts of perturbations. In this paper, we develop a class of linearly solvable risk-sensitive problems. The degree of risk-sensitivity is controlled by a scalar parameter $\alpha$. For $\alpha < 0$, we get risk-seeking behavior and for $\alpha > 0$, we get risk-averse behavior. We show that the Bellman equation characterizing the optimal solution to problems in this class can be made linear (using an exponential transformation) and call these problems

$\alpha$-Risk Sensitive Linearly Solvable Control Problems ($\alpha$-**RLCs**).

Our results can be seen as a generalization of 2 lines of work: one on Linearly Solvable MDPs(**LMDPs**) (Todorov, 2009) and the other on risk-sensitive path-integral control (Broek *et al.*, 2010). We obtain the **LMDP** results in the risk-neutral limit $\alpha \to 0$ and the results from (Broek *et al.*, 2010) by taking a continuous-time limit. To the best of our knowledge, $\alpha$-**RLCs** are the broadest class of linearly solvable control problems known and include all previous work as a special case.

This paper is organized as follows: Section 2 introduces notation and preliminaries. Section 3 contains the main $\alpha$-**RLC** results and discusses connections between $\alpha$-**RLCs** and previous work. Theorem 1 is the main technical result of the paper, that uses an elegant result on $\alpha$ divergences proved in Theorem 5. Section 3.4 discusses interesting properties of $\alpha$-**RLCs**: Path Integral Representations of the optimal value function (section 3.4.2), Compositionality of Optimal Control Laws (section 3.4.1) and a Game Theoretic Interpretation of $\alpha$-**RLCs**(section 3.1). In Section 4, we present numerical results on simple control problems that illustrate the risk-sensitive behavior.

## 2 BACKGROUND

### 2.1 NOTATION

We use $\mathcal{X}$ to denote the state space, $x$ to denote states. We assume that $\mathcal{X}$ is a finite set. $v^{(\alpha)}$ for the optimal value function under risk-parameter $\alpha$. Let $\mathcal{U}_c(x)$ denote the space of feasible control signals in state $x$ and $\mathcal{P}[\mathcal{X}]$ be the set of probability distributions over $\mathcal{X}$. We use $u_c$ to denote an arbitrary member of $\mathcal{U}_c(x)$. For any $p \in \mathcal{P}[\mathcal{X}]$, let $\text{supp}[p] = \{x \in \mathcal{X} : p(x) > 0\}$. We use $\pi_{co}$ to denote policies, ie, mappings from states to controls, $\pi_{co}(x) \in \mathcal{U}_c(x)$ for every $x$. In this paper, controls will be probability distributions over $\mathcal{X}$ themselves, and we use the notation $\pi_{co}(\cdot|x)$ to denote the control distribution in state $x$. We use $v^{(\alpha)}(\cdot, \pi_{co})$ to denote the policy-specific value function at risk factor $\alpha$. Define the KL-divergence between two members of $\mathcal{P}[\mathcal{X}]$ by $\text{KL}(\pi_{co} \| \pi_0) = \sum_{x \in \mathcal{X}} \pi_{co}(x) \log\left(\frac{\pi_{co}(x)}{\pi_0(x)}\right)$, which is well-defined when $\text{supp}[\pi_{co}] \subseteq \text{supp}[\pi_0]$. Let $f$ be any real-valued function on $\mathcal{X}$. We denote the expectation of $f$ under $\pi$ as $\mathrm{E}_\pi[f] = \sum_x \pi(x) f(x)$ and let $\Psi^\alpha_\pi[f] = \alpha^{-1} \log\left(\mathrm{E}_\pi[\exp(\alpha f)]\right)$ and $\Psi_\pi[f] = \Psi^1_\pi[f]$. One can prove that in the limit $\alpha \to 0$, this is just the expectation so we define $\Psi^0_\pi[f] = \mathrm{E}_\pi[f]$. Note that both these quantities are independent of $x$. We denote a Gaussian distribution as $\mathcal{N}(\mu, \Sigma)$ and the corresponding density value at a point $x$ as $\mathcal{N}(x; \mu, \Sigma)$.

### 2.2 RISK SENSITIVE MDPs

Risk Sensitive MDPs (Marcus *et al.*, 1997) are defined by specifying a state space $\mathcal{X}$, a control space $\mathcal{U}_c(x)$ for each $x \in \mathcal{X}$, a transition probability function $P$ ($P(x'|x, u)$ is the probability of landing in $x' \in \mathcal{X}$ in one time step after applying control $u$ is state $x$), a cost function $c(x, u)$ and a risk-sensitivity parameter $\alpha$. Let $P_\pi$ denote the dynamics of the sytems under control policy $\pi$. The objective is to design a policy $\pi_{co}(x) \in \mathcal{U}_c(x)$ that minimizes an accumulated cost, which is defined in the finite horizon(FH),infinite horizon(IH) and first-exit(FE) problems as follows:

$$\text{FH}: \Psi^\alpha_{P_{\pi_{co}}}\left[\exp\left(\alpha \sum_{t=0}^T c(x_t, \pi_{co}(x_t), t)\right)\right]$$

$$\text{IH}: \lim_{T \to \infty} \frac{1}{T} \Psi^\alpha_{P_{\pi_{co}}}\left[\exp\left(\alpha \sum_{t=0}^T c(x_t, \pi_{co}(x_t))\right)\right]$$

$$\text{FE}: \Psi^\alpha_{P_{\pi_{co}}}\left[\exp\left(\alpha \sum_{t=0}^{T_e} c(x_t, \pi_{co}(x_t))\right)\right]$$

where $T_e$ denotes the first time the system enters a *terminal state*. A first order approximation of the FH objective gives $\mathrm{E}_{P_{\pi_{co}}}\left[\sum_{t=0}^T c(x_t, \pi_{co}(x_t))\right] + \alpha \text{Var}\left(\sum_t c(x_t, \pi_{co}(x_t))\right)$. Thus, if $\alpha > 0$, the cost increases with variance of the accumulated cost while if $\alpha < 0$, the cost decreases with variance so that we get risk-seeking behavior. As $\alpha \to 0$, we get the standard MDP back. Thus, one can effectively trade risk and expected costs: If one is willing to take risks (increase variance of costs), it is possible to get achieve lower expected cost. On the other hand, if one wants strong guarantees on expected cost (low variance), then one needs to settle for a higher expected cost. The degree of risk-averse/risk-seeking behavior increases with the magnitude of $\alpha$. The discounted problem in risk-sensitive MDPs does not admit a stationary optimal policy in general and hence we do not consider it here. It can be shown that the Bellman equation(Marcus *et al.*, 1997) for risk-sensitive MDPs is given by

$$\text{FH}: v^{(\alpha)}_t(x) = \min_{u_c \in \mathcal{U}_c(x)} c(x, u_c, t) + \Psi^\alpha_{P_\pi(\cdot|x)}\left[v^{(\alpha)}_{t+1}\right]$$

$$\text{IH}: v^{(\alpha)}(x) + \bar{c} = \min_{u_c \in \mathcal{U}_c(x)} c(x, u_c) + \Psi^\alpha_{P_\pi(\cdot|x)}\left[v^{(\alpha)}\right]$$

$$\text{FE}: v^{(\alpha)}(x) = \min_{u_c \in \mathcal{U}_c(x)} c(x, u_c) + \Psi^\alpha_{P_\pi(\cdot|x)}\left[v^{(\alpha)}\right] \quad (1)$$

where $\bar{c}$ is the optimal average risk-sensitive cost in the IH formulation. For FH problems, the value function at the end of the horizon is given by the final cost: $v^{(\alpha)}_T(x) = c(x, T)$. In FE problems the value function on the exit manifold $\mathcal{T}$ is given by the final cost:$v^{(\alpha)}(x) = c_f(x)$ for $x \in \mathcal{T}$ in FE problems. Exis-tence of solutions to (1) in continuous state spaces has been a subject of active research but we do not discuss it in detail in this paper. Recent papers (Jaśkiewicz, 2007) have shown the existence of solutions to the IH Bellman equation under fairly general conditions.

## 2.3 RÉNYI DIVERGENCES

Rényi divergences (Van Erven & Harremoes, 2010) are a general class of divergences between probability distributions that generalize the well-known KL-divergence.

**Definition 1.** Let $p, q \in \mathcal{P}[\mathcal{X}]$. Then the Rényi divergence between $p, q$ of order $\alpha$ is defined as $\mathbb{D}_\alpha(p \parallel q) = \frac{1}{\alpha(\alpha-1)} \log \left( \sum_{x \in \mathcal{X}} p(x)^\alpha q(x)^{1-\alpha} \right)$ if $\alpha \neq 0, 1$ and $\mathbb{D}_0(p \parallel q) = \mathrm{KL}(q \parallel p), \mathbb{D}_1(p \parallel q) = (p \parallel q) = \mathrm{KL}(p \parallel q)$. The Rényi divergence is well defined if $\mathrm{supp}[p] \cap \mathrm{supp}[q] \neq \emptyset$ if $\alpha \in (0,1)$, $\mathrm{supp}[p] \subset \mathrm{supp}[q]$ if $\alpha \geq 1$ and $\mathrm{supp}[q] \subset \mathrm{supp}[p]$ if $\alpha \leq 0$. It can be shown (lemma 1) that $\mathbb{D}_\alpha(p \parallel q) \geq 0, (\mathbb{D}_\alpha(p \parallel q) = 0 \iff p = q)$ for any $\alpha \in \mathbf{R}$.

This definition differs from the standard definition by a factor of $\alpha$ (the standard definition does not cover the case $\alpha \leq 0$ either), but the modified definition is more convenient to work with. We defined the limiting cases $\mathbb{D}_0, \mathbb{D}_1$ by taking limits as $\alpha \to 0, \alpha \to 1$ so that $\mathbb{D}_\alpha$ is a continuous function of $\alpha$.

## 2.4 LMDPs

Linearly solvable MDPs (Todorov, 2009) are a class of MDPs for which the Bellman equation characterizing the optimal value function can be made linear. In traditional MDPs, the controller chooses an action $u \in \mathcal{U}_c$ that then determines the probability of the future state through the transition density $P(x'|x, u)$. In **LMDPs**, we allow the controller to directly choose the transition density $\pi_{co}(x'|x)$, that is, $\mathcal{U}_c = \mathcal{P}[\mathcal{X}]$. However, if no restrictions were placed on this, the optimal $\pi_{co}(x'|x)$ could be trivially chosen to be a delta-distribution centered at the lowest-cost state. In order to get an interesting problem, we impose a cost on picking a transition density: this is defined as the KL-divergence between $\pi_{co}$ and $\pi_0$, the "passive" dynamics of the system, ie, the dynamics of the system left to itself (when the controller does nothing). Thus, we can think of the controller as trying to change the natural dynamics of the system in order to move towards states of low cost, but it pays a price for doing so. Thus, the total cost function is given by $c(x, u_c, t) = q(x) + \mathrm{KL}(u_c \parallel \pi_0(\cdot|x))$ where $q(x)$ is any cost on the state. It can be shown (Todorov, 2009) that the exponentiated optimal value function

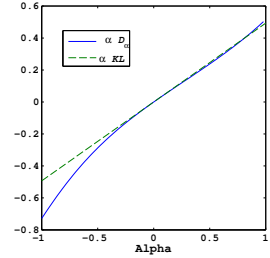

Figure 1: $\alpha \mathbb{D}_\alpha(p \parallel q)$ vs $\alpha \mathrm{KL}(q \parallel p)$

$z_t(x) = \exp(-v_t(x))$ satisfies a linear Bellman equation: $z_t(x) = \exp(-q(x,t)) \sum_{x' \in \mathcal{X}} \pi_0(x'|x) z_t(x')$.

## 3 LINEARLY SOLVABLE RISK SENSITIVE CONTROL

In this section, we show that it is possible to extend the framework of **LMDPs** to Risk Sensitive MDPs while preserving the linearity of the Bellman equation. However, we need to modify the risk sensitive formulation slightly in order to do this.

**Definition 2.** A Linearly Solvable Risk Sensitive Control Problem ($\alpha$-**RLC**) is a Risk-Sensitive MDP with risk parameter $\alpha$ and the following restrictions:

The actions are the controlled dynamics themslves: $u_c \in \mathcal{U}_c(x) \subset \mathcal{P}[\mathcal{X}], P(x'|x, u_c) = u_c(x')$.

The basic cost function can be decomposed as follows: $c(x, u_c) = q(x) + \mathrm{KL}(u_c \parallel \pi_0(\cdot|x))$ where $\pi_0$ is called the *Passive Dynamics* and corresponds to the natural dynamics of the system in the absence of controls. However, at risk factor $\alpha$, we replace the scaled cost $\alpha c$ with $\alpha q(x) + \alpha \mathbb{D}_\alpha(\pi_0(\cdot|x) \parallel u_c)$. Thus, the FH objective (scaled by $\alpha$) becomes :

$$\log \left( \mathrm{E}_{\pi_{co}} \exp \left( \alpha \sum_{t=0}^T q(x_t) + \mathbb{D}_\alpha(\pi_0(\cdot|x_t) \parallel \pi_{co}(\cdot|x_t)) \right) \right)$$

We have $\mathcal{U}_c(x) = \{u_c \in \mathcal{P}[\mathcal{X}] : \mathrm{supp}[u_c] \subset \mathrm{supp}[\pi_0(\cdot|x)]\}$ if $\alpha \leq 0$, $\mathcal{U}_c(x) = \{u_c \in \mathcal{P}[\mathcal{X}] : \mathrm{supp}[u_c] \supset \mathrm{supp}[\pi_0(\cdot|x)]\}$ if $\alpha \geq 1$ and $\mathcal{U}_c(x) = \{u_c \in \mathcal{P}[\mathcal{X}] : \mathrm{supp}[u_c] \cap \mathrm{supp}[\pi_0(\cdot|x)] \neq \emptyset\}$ if $0 < \alpha < 1$.

If the regular risk-sensitive framework is applied to **LMDPs**, we would simply exponentiate the scaled cost:

$$\exp \left( \sum_{t=0}^T \alpha q(x_t) + \alpha \mathrm{KL}(\pi_{co}(\cdot|x_t) \parallel \pi_0(\cdot|x_t)) \right)$$

However, in this alternate approach, we replace $\alpha \mathrm{KL}(\pi_{co}(\cdot|x_t) \parallel \pi_0(\cdot|x_t))$ by the Rényi divergence $\alpha \mathbb{D}_\alpha(\pi_{co}(\cdot|x_t) \parallel \pi_0(\cdot|x_t))$. Even though it differs from

the standard formulation, this is justified by the following observations:

1) As $\alpha \to 0, \mathbb{D}_\alpha \to$ KL (lemma 1). Thus, in the limit $\alpha \to 0$, we recover the risk-neutral LMDP

2) $\alpha \mathbb{D}_\alpha$ is a monotonically increasing function of $\alpha$ (lemma 1), negative for $\alpha < 0$ and positive for $\alpha > 0$ just like the linear scaling.

3) If $p, q$ are Gaussians with the same covariance, $\alpha \mathbb{D}_\alpha (p \parallel q) = \alpha \text{KL} (q \parallel p)$ (lemma 2) so we're doing the traditional linear scaling.

4) The game theoretic interpretation (section 3.1) of this framework further demonstrates how it naturally models to risk sensitivity. For risk averse problems, the game theoretic interpretation says that the controller is fighting against a stronger adversary as $\alpha$ increases, and hence is expected to be more conservative/risk-averse while the reverse is true as $\alpha$ becomes negative and large.

5) Numerical experiments (section 4) that show that this new risk sensitive model conforms to intuitive notions of risk.

For two randomly generated distributions $p, q$, the linear and nonlinear scaling are plotted in figure 1.

**Theorem 1.** *The Bellman equation for an $\alpha$-**RLC** can be made linear in a transformed value function. If $\alpha \neq 1, z^{(\alpha)}(x) = \exp\left(-(1-\alpha)v^{(\alpha)}(x)\right), Q(x) = \exp\left(-(1-\alpha)q(x)\right)$. The resulting linear bellman equation is, if $\alpha \neq 1$:*

FH:     $z_t^{(\alpha)}(x) = Q(x,t) \operatorname{E}_{\pi_0(\cdot|x)} \left[z_{t+1}^{(\alpha)}\right]$

FE:     $z^{(\alpha)}(x) = Q(x) \operatorname{E}_{\pi_0(\cdot|x)} \left[z^{(\alpha)}\right]$

IH:     $z^{(\alpha)}(x) = Q(x) \exp\left(-(\alpha-1)\bar{c}\right) \operatorname{E}_{\pi_0(\cdot|x)} \left[z^{(\alpha)}\right]$

FH:     $v_t^{(1)}(x) = q(x,t) + \operatorname{E}_{\pi_0(\cdot|x)} \left[v_{t+1}^{(1)}\right]$

FE:     $v^{(1)}(x) = q(x) + \operatorname{E}_{\pi_0(\cdot|x)} \left[v^{(1)}\right]$

IH:   $v^{(1)}(x) + \bar{c} = q(x) + \operatorname{E}_{\pi_0(\cdot|x)} \left[v^{(1)}\right]$  (2)

*For FH problems, $v_T^{(\alpha)}(x) = q(x,T)$. For FE problems, $\mathcal{T}, \mathcal{N}$ are the terminal and non-terminal states, respectively and $v^{(\alpha)}(x) = q(x) \forall x \in \mathcal{T}$. For IH problems, $\bar{c}$ is the optimal average cost for infinite-horizon average cost problems. The optimal control law is always given by $\pi_{co}^*(x'|x) = \frac{\pi_0(x'|x)\exp\left(-v^{(\alpha)}(x')\right)}{\int \pi_0(x'|x)\exp\left(-v^{(\alpha)}(x')\right) dx'}$.*

*Proof.* We prove the result for the IH case, the proof for the other cases is similar. The Bellman equation (1) becomes
$v^{(\alpha)}(x) + \bar{c} =$
$q(x) + \min_{u_c \in \mathcal{U}_c(x)} \mathbb{D}_\alpha \left(\pi_0(\cdot|x) \parallel u_c\right) + \Psi_{u(\cdot|x)}^\alpha \left[v^{(\alpha)}\right] = q(x) + \Psi_{\pi_0(\cdot|x)}^{\alpha-1} \left[v^{(\alpha)}\right]$   (Using theorem 5)
If $\alpha = 1$, this is already a linear Bellman equation since $\Psi_{\pi_0(\cdot|x)}^0 \left[v^{(\alpha)}\right] = \operatorname{E}_{\pi_0(\cdot|x)} \left[v^{(\alpha)}\right]$. If not, we have:
$(\alpha - 1)(v^{(\alpha)}(x) + \bar{c} - q(x,t)) = \Psi_{\pi_0(\cdot|x)} \left[(\alpha-1)v^{(\alpha)}\right]$
Exponentiating both sides, we get:
$z^{(\alpha)}(x) = \exp\left(-(\alpha-1)\bar{c}\right) Q(x) \operatorname{E}_{\pi_0(\cdot|x)} \left[z^{(\alpha)}\right]$   □

### 3.1 GAME-THEORETIC INTERPRETATION

The relationship between risk-sensitive control and dynamic games has been studied extensively (Fleming & McEneaney, 1992). For systems with linear dynamics, quadratic costs and Gaussian noise (LQG systems), the classic paper of Jacobson et al (Jacobson, 1973) showed exact equivalence between risk-sensitive control and deterministic LQ games (2 player zero-sum games with deterministic linear dynamics and quadratic costs). We generalize that result here and show exact correspondence between $\alpha$-**RLCs** and a class of 2-player zero-sum Markov Games. If $\alpha > 0$, $\alpha$-**RLCs** can be interpreted as solving a 2-player zero-sum Markov Game (Başar & Bernhard, 1995), where the first player (controller) is trying to minimize the cost while the second player (adversary) is trying to maximize it. We have analogous results if $\alpha < 0$ and one considers a cooperative (rather than competitive) game where both players cooperate to minimize costs. First suppose $\alpha > 0$(the construction/proof for the other case is similar). Consider a Markov game that proceeds as follows:
System is in state $x$
Player 1 (controller) picks transition density $\pi_{co}(\cdot|x)$
Player 2 (adversary) picks transition density $\pi_a(\cdot|x, \pi_{co})$
The system transitions to state $x' \sim \pi_a(\cdot|x, \pi_{co})$
The dynamics of the system is completely controlled by the adversary and the controller has no direct influence on it. However, the adversary pays a cost for picking a transition drastically different from the controller: $c(x, \pi_{co}(\cdot|x), \pi_a(\cdot|x), t) = q(x,t) + \mathbb{D}_\alpha \left(\pi_{co}(\cdot|x) \parallel \pi_0(\cdot|x)\right) - \frac{1}{\alpha} \text{KL}\left(\pi_a(\cdot|x) \parallel \pi_{co}(\cdot|x)\right)$.
The **upper value** function (Başar & Bernhard, 1995) of the game is given by the Bellman-Issacs equation(Başar & Bernhard, 1995):
$v_t^{(\alpha)}(x) = \min_{u_c \in \mathcal{U}_c} \max_{u_a \in \mathcal{U}_a} c(x, u_c, u_a) + \operatorname{E}_{u_a}\left[v_{t+1}^{(\alpha)}\right]$
$v_t^{(\alpha)}(x) = q(x,t) + \min_{u_c \in \mathcal{U}_c} \mathbb{D}_\alpha \left(u_c \parallel \pi_0(\cdot|x)\right) - \min_{u_a \in \mathcal{U}_a} \frac{\text{KL}\left(u_a \parallel u_c\right)}{\alpha} - \operatorname{E}_{u_a} v^{(\alpha)}$
$v_t^{(\alpha)}(x) = q(x,t) + \min_{u_c} \mathbb{D}_\alpha \left(u_c \parallel \pi_0(\cdot|x)\right) + \Psi_{u_c}^\alpha \left[v^{(\alpha)}\right]$
where the last line follows from using theorem 5 in the limit $\alpha \to 0$. The last equation matches the risk-sensitive Bellman equation (1) and hence the optimal strategy for the controller matches that for the $\alpha$-**RLC**. The optimal strategy for the adversary (using theorem 5) is

$$\pi_a^*(x'|x,t) \propto \pi_{co}^*(x'|x,t) \exp\left(\alpha v_{t+1}^{(\alpha)}(x')\right)$$
$$\propto \pi_0(x'|x) \exp\left((\alpha-1)v_{t+1}^{(\alpha)}(x')\right)$$

If we scale the control costs of both the controller and the adversary by $\alpha$, we see that the controller's cost $\alpha \mathbb{D}_\alpha$ increases monotonically while the adversary's cost KL remains fixed. Thus, as $\alpha$ increases, the controller is fighting against a stronger adversary and will tend to be more conservative/risk-averse. This further justifies the nonlinear scaling used in the risk-sensitive interpretation.

In the game theoretic setting, if $\alpha > 1$, the closed loop system (which mimics the adversary) has a dynamics that moves towards increasing costs (since the cost-to-go $v^{(\alpha)}$ has a positive coefficient), meaning that the adversary dominates the game. If $0 < \alpha < 1$, on the other hand, the closed loop system has a dynamics that moves towards decreasing costs and the controller dominates the game.

## 3.2 EXISTENCE OF SOLUTIONS

In this section, we show that under reasonable assumptions, (2) has solutions when $\mathcal{X}$ is finite. The extension to infinite $\mathcal{X}$ should be possible under appropriate technical assumptions, but we leave that for future work.

**Theorem 2.** Let $\mathcal{X}$ be finite, $|\mathcal{X}| = \text{ns}$. Let $z \in \mathbf{R}^{\text{ns}} = \{z^{(\alpha)}(x) : x \in \mathcal{X}\}$ and similarly define $v^{(\alpha)}, q \in \mathbf{R}^{\text{ns}}$ and $\pi_0 \in \mathbf{R}^{\text{ns} \times \text{ns}}$. (2) always has a solution in FH. Suppose that $\pi_0$ is irreducible, ie, there is a path of non-zero probability from every state to every other state under the passive dynamics. Then, (2) has a solution in IH. Further, if $q \geq 0, \alpha \leq 1$, (2) has a solution in FE.

*Proof.* Solutions to the Bellman equations (2) always exist in the finite horizon case (Marcus *et al.*, 1997). The Bellman equation (2) for the IH case becomes $z^{(\alpha)} = \exp\left(-(\alpha-1)\bar{c}\right) \text{diag}\left(\exp\left((\alpha-1)q\right)\right) \pi_0 z^{(\alpha)}$ if $\alpha \neq 1$ and $v^{(\alpha)} + \bar{c} = q + \pi_0 v^{(\alpha)}$ if $\alpha = 1$. In the first case, we have an eigenvalue problem that is guaranteed to have a unique positive eigenvector under the Perron-Frobenius theorem (Serre, 2010) given that $\text{diag}\left(\exp\left((\alpha-1)q\right)\right) \pi_0$ is irreducible. If $\alpha = 1$, we can enforce the extra constraint $\sum v^{(\alpha)}(x) = 0$ to get linear system in $v^{(\alpha)}, \bar{c}$ which is non-singular (and hence has a unique solution) as long as $\pi_0$ is irreducible. In the FE case, if $\alpha = 1$, we can break up the equation into terminal and non-terminal parts

$$(I - \text{diag}\left(Q^\mathcal{N}\right)\pi_0^{\mathcal{N}\mathcal{N}})z^{(\alpha)\mathcal{N}} = \pi_0^{\mathcal{N}\mathcal{T}}Q^{\mathcal{T}}$$

where the superscripts indicate indexing, $Q = \exp\left((\alpha-1)q\right)$. If $\alpha < 1, q \geq 0$, $\text{diag}\left(Q^\mathcal{N}\right)\pi_0^{\mathcal{N}\mathcal{N}}$ has a spectral radius less than 1 so that the above system has a positive solution (Serre, 2010). Again, the case $\alpha = 1$ is easy to handle since its directly linear in $v^{(\alpha)}$ space. $\square$

## 3.3 CONNECTIONS TO PREVIOUS WORK

In this section, we draw connections between $\alpha$-**RLCs** and previous work: In particular **LMDPs** (Todorov, 2009), Risk-Sensitive Path Integral Control (Broek *et al.*, 2010) and policy iteration.

### 3.3.1 LMDPs

As $\alpha \to 0$, in the limit $\mathbb{D}_\alpha(p \parallel q) \to \text{KL}(q \parallel p)$. Thus, in the limit $\alpha \to 0$, $\alpha$-**RLC** becomes equivalent to an MDP with cost function $q(x) + \text{KL}(\pi_0(\cdot|x) \parallel u)$. The Bellman equation (2) becomes $z^{(0)}(x) = \exp\left(-v^{(0)}(x)\right), z^{(0)}(x) = \exp\left(\bar{c} - q(x)\right) E_{\pi_0(\cdot|x)}\left[z^{(0)}\right]$ which is exactly the **LMDP** Bellman equation (Todorov, 2009). Thus, **LMDPs** are a special case of $\alpha$-**RLCs** obtained in the limit when $\alpha \to 0$. We shall see in section 3.4 that many of the nice properties of **LMDPs** generalize to $\alpha$-**RLCs**. The policy gradient theorem and the maximum principle (Todorov, 2009), however, do not generalize to $\alpha$-**RLCs**.

### 3.3.2 Risk Sensitive Path-Integral Control

Suppose we have an Ito diffusion process:

$$dx = a(x)dt + B(x)(udt + \sigma d\omega)$$

with a cost rate $c(x,u) = q(x) + \frac{\lambda}{2\sigma^2} u^T u$. Consider an $h$-step Euler discretization of the problem:
$P^h(x'|x,u) = \mathcal{N}(x + (a(x) + B(x)u)h; \sigma h B(x)B(x)^T)$
Let $z_h^{(\alpha)}(x)$ be the $z$ function for the h-step discretization and $\pi_0^h(x'|x) = \mathcal{N}(x + a(x)h; \sigma h B(x)B(x)^T)$. The Bellman equation for the $\alpha$-**RLC** with $\alpha' = \lambda\alpha, \beta = 1 - \alpha'$ is

$$z_h^{(\alpha)}(x) \exp\left(h\beta(q(x) - \bar{c})\right) = E_{\pi_0^h(x'|x)}\left[z_h^{(\alpha)}(x')\right]$$

$$z_h^{(\alpha)}(x) \frac{\exp\left(h\beta(q(x) - \bar{c})\right) - 1}{h} = E\left[\frac{z_h^{(\alpha)}(x') - z_h^{(\alpha)}(x)}{h}\right]$$

Taking limits as $h \to 0$, we know from stochastic calculus (Øksendal, 2003) that the RHS becomes the generator of the passive Ito diffusion $dx = a(x)dt + \sigma B(x)d\omega$ applied to $z^{(\alpha)}$:

$$z^{(\alpha)}(x)(q(x) - \bar{c}) = \frac{\nabla z^{(\alpha)}(x)^T a(x) + \frac{\text{tr}\left(B(x)B(x)^T \nabla^2 z(x)\right)}{2}}{1 - \lambda\alpha}$$

Also, using lemma 2, we get $\alpha' \mathbb{D}_\alpha\left(\pi_0^h(\cdot|x) \parallel P^h(\cdot|x,u)\right) = \frac{h\alpha' u^T u}{2\sigma^2} = \frac{h\lambda\alpha u^T u}{2\sigma^2}$

so that $c_\alpha^h(x, P(\cdot|x,u)) = \alpha h c(x,u)$. Thus, we're solving exactly the traditional risk sensitive control problem for the given Ito process as $h \to 0$. These results also coincide with those in (Broek *et al.*, 2010), so Risk-Sensitive Path Integral Control can be seen as a continuous-time limit of $\alpha$-**RLCs**.

### 3.3.3 Policy Iteration

There is a very interesting relationship between $\alpha$-**RLCs** with $\alpha = \theta$ and $\alpha = \theta - 1$. The optimal value function for $\alpha = \theta$ is given by $\mathrm{v}_t^{(\theta)}(x) = q(x) + \Psi_{\pi_0(\cdot|x)}^{\theta-1}\left[\mathrm{v}_{t+1}^{(\theta)}\right]$. The value function corresponding to the policy $\pi$ when $\alpha = \theta - 1$ is given by solving the policy-specific Bellman equation: $\mathrm{v}_t^{(\theta-1,\pi)}(x) = q(x) + \mathbb{D}_{\theta-1}\left(\pi_0(\cdot|x) \parallel \pi(\cdot|x)\right) + \Psi_\pi^{\theta-1}\left[\mathrm{v}_{t+1}^{(\theta-1,\pi)}\right]$. If we plug in $\pi = \pi_0$, this becomes identical to the previous equation. Hence, the optimal value function for $\alpha = \theta$ corresponds to the value of the null policy $\pi_0$ with $\alpha = \theta - 1$. Thus, if we run policy iteration when $\alpha = \theta - 1$ starting with the null policy $\pi_0$, in one step we get
$\mathrm{argmin}_{u_c \in \mathcal{U}_c} \mathbb{D}_{\theta-1}\left(\pi_0(\cdot|x) \parallel u_c\right) + \Psi_\pi^{\theta-1}\left[\mathrm{v}^{(\theta)}\right] = \pi_{co}^*(\theta|x)$ using lemma 5. Thus, the optimal policy for risk parameter $\theta$ is the policy obtained after one step of policy iteration with risk parameter $\theta - 1$.

### 3.4 PROPERTIES OF $\alpha$-RLCs

In this section, we show that for $\alpha$-**RLCs**, under some conditions, it is possible to combine the solutions of simple control problems directly to solve a more complicated control problem. This property has been used in the risk neutral case for building controllers for character animation (Da Silva *et al.*, 2009).

#### 3.4.1 Compositionality

**Theorem 3.** *Let $\{q_f^i\}_{i=1}^k$ be a set of final costs for a set of $\alpha$-RLC FH or FE problems with the same passive dynamics $\pi_0$ and running cost $q(x)$ and let $w \in \mathbf{R}^k, w \geq 0$. If $\alpha \neq 1$, let $\{\mathrm{z}_i^{(\alpha)}\}$ be the set of optimal z-functions corresponding to each final cost. Then, $\sum_i w_i \mathrm{z}_i^{(\alpha)}$ is the optimal z-function for the problem with final cost $\frac{1}{\alpha-1}\log\left(\sum_i w_i \exp\left((\alpha-1)q_f^i\right)\right)$. If $\alpha = 1$, the $\sum_i w_i \mathrm{v}_i^{(\alpha)}$ is the optimal value function for the composite final cost $\sum_i w_i q_f^i$. In this case, both running and final costs can be composed.*

*Proof.* We do the proof for FE the FH case is similar We know that for each $i$, we have
$\mathrm{z}_i^{(\alpha)}(x) = \exp\left((\alpha-1)q(x)\right)\mathrm{E}_{\pi_0}\left[\mathrm{z}_i^{(\alpha)}\right]$ if $x \in \mathcal{N}$
$\mathrm{z}_i^{(\alpha)}(x) = \exp\left((\alpha-1)q_f^i(x)\right)$ if $x \in \mathcal{T}$
Thus, $\mathrm{z}^{(\alpha)}(x) = \sum_i w_i \mathrm{z}_i^{(\alpha)}(x)$, by linearity, must satisfy
$\mathrm{z}^{(\alpha)}(x) = \exp\left((\alpha-1)q(x)\right)\mathrm{E}_{\pi_0}\left[\mathrm{z}^{(\alpha)}\right]$ if $x \in \mathcal{N}$
$\mathrm{z}^{(\alpha)}(x) = \sum_i w_i \exp\left((\alpha-1)q_f^i(x)\right)$ if $x \in \mathcal{T}$
which means that $\mathrm{z}^{(\alpha)}$ is the optimal $z$-function for the problem with composite cost $q_f(x) = \frac{1}{\alpha-1}\log\left(\sum_i w_i \exp\left((\alpha-1)q_f^i\right)\right)$. For $\alpha = 1$, (2) is linear in $\mathrm{v}^{(\alpha)}, q$, so both running and final costs can be composed. □

#### 3.4.2 Path-Integral Representation

In this section, we show that the optimal value function for $\alpha$-**RLCs** has a path integral representation, ie, it can be expressed as an expectation of the transformed cost over trajectories sampled from the passive (or uncontrolled) dynamics. For the risk-neutral case, this property has been used to build controllers for practical robotic applications successfully (Theodorou *et al.*, 2010).

**Theorem 4.** *If $\alpha \neq 1$, the optimal value function has a path integral representation:*

$$\mathrm{FH}: \mathrm{v}_0^{(\alpha)}(x) = \frac{\log\left(\mathrm{E}_{\pi_0}\left[\exp\left((\alpha-1)q(x_{0:T})\right)|x_0 = x\right]\right)}{\alpha - 1}$$

$$\mathrm{FE}: \mathrm{v}^{(\alpha)}(x) = \frac{\log\left(\mathrm{E}_{\pi_0}\left[\exp\left((\alpha-1)q(x_{0:T_e})\right)|x_0 = x\right]\right)}{\alpha - 1}$$
(3)

*where the expectation is over trajectories starting from $x$ sampled under the passive dynamics $\pi_0$ and $q(x_{0:\tau}) = \sum_{t=0}^\tau q(x_t)$. $T_e$ is the first time the trajectory enters a terminal state $\mathcal{T}$. When $\alpha = 1$, we have $\mathrm{v}_0^{(\alpha)}(x) = \mathrm{E}_{\pi_0}[q(x_{0:T})|x_0 = x]$ for FH and $\mathrm{v}^{(\alpha)}(x) = \mathrm{E}_{\pi_0(\cdot|x)}[q(x_{0:T_e})|x_0 = x]$ for FE.*

*Proof.* From section 3.3.3, we know that $\mathrm{v}_0^{(\alpha)}(x) = \mathrm{v}_0^{(\alpha-1)}(x, \pi_0)$. The RHS of (3) is precisely the value of the policy $\pi_0$ under risk $\alpha$. □

## 4 EXPERIMENTS

We illustrate the effects of the risk-sensitivity parameter and the uses of a risk-sensitive objective through a control problem. The state space is $[-3, 3] \times [-6, 6]$ and encodes the position and velocity of a point mass moving along a hilly terrain. The equation describing the terrain is given by:
$f(x) = \exp\left(-\frac{v_1(x-0.9)^2}{2}\right) + r\exp\left(-\frac{v_2(x+0.9)^2}{2}\right)$
ie, a superposition of 2 hills peaked at $x = -0.9$ and

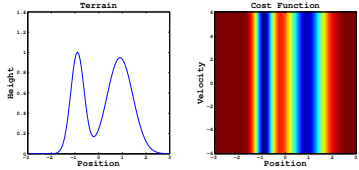

Figure 2: Terrain and State Cost (blue is low cost and red high)

$x = 0.9$ respectively. The uncontrolled (passive) dynamics of the system are given by the effect of gravity (acting vertically downward) on the car moving along the surface (so that only the component of gravity along the tangent to the surface matters):

$dp = \frac{v}{\sqrt{1+f'(p)^2}} dt, dv = -\frac{gf'(p)}{\sqrt{1+f'(p)^2}} dt + \sigma d\omega$

We choose $r = 0.95, v_1 = 12.5, v_2 = 3.4, \sigma = 2$ for our experiments. We consider the $h$-step Euler discretization of the problem to get a passive dynamics $\pi_0$ and define a state cost $q(x) = 1 - f(x)$ encouraging the point mass to stay near the peaks. The terrain and state costs are shown in figure 2. In order to solve the problem, we discretize the state space $\mathcal{X}$ with a $101 \times 101$ grid. We can the define $q, \pi_0$ on the grid and solve the IH problem using the power-iteration method, which works fairly well since $\pi_0$ is sparse. Using a straightforward implementation in MATLAB, it takes about 0.5 seconds to solve the problem.

We solve this problem for 3 different values of alpha $-0.1$(risk-seeking),$0$(risk-neutral) and $0.1$(risk-averse). We plot the stationary distribution of the optimally controlled system for each case in figure 3. As expected, the stationary distribution is concentrated around the 2 peaks in all cases. However, since the hill peaked at $+0.9$ is slightly less steep (albeit shorter), it is less risky (a small perturbation is less likely to push one downhill to high cost regions). Thus, the risk-averse controller sticks to the shorter/broader hill. The risk-seeking controller chooses the lower cost but riskier taller hill. The risk-neutral controller, places almost equal probability on both solutions. Thus, we can see that the risk-sensitive framework allows us to trade risk for reward (or low cost).

## 5 CONCLUSIONS

We have developed a very general family of linearly solvable control problems. To the best of our knowledge, all previous work on linearly solvable control are special cases. Also, the use of Rényi divergences in control is novel. An interesting theoretical question is whether $\alpha$-**RLCs** are the most general family of linearly solvable control problems possible.

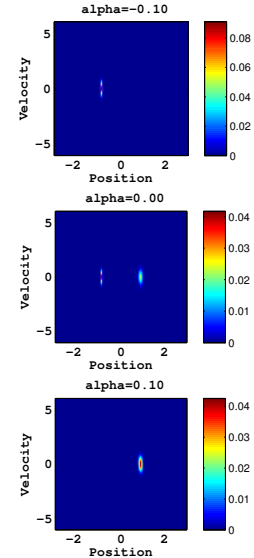

Figure 3: Optimal Stationary Distribution vs $\alpha$

In terms of practical applicability, $\alpha$-**RLCs** could be very useful for tuning controllers to be more conservative (risk-averse) or more aggressive (risk-taking). We have seen that the resulting behavior can be substantially different for different $\alpha$. The linearity makes $\alpha$-**RLCs** easier to solve, but we still need to develop function approximation techniques that scale to high dimensional state spaces and nonlinear dynamics. If $\pi_0$ is Gaussian and $z^{(\alpha)}$ is represented as a mixtures of Gaussians $\times$ polynomials $\times$ trigonometric functions, one can use a power-iteration like algorithm to solve the linear Bellman equation with each step being analytical. Combined with the flexibility of $\alpha$-**RLCs** we believe that these techniques are very promising and can potentially solve hard control problems.

## 6 Appendix

To avoid technical complications, we do the proof only for the case when $\mathcal{X}$ is finite. Let $\pi_0 \in \mathcal{P}[\mathcal{X}]$, $f$ be any real-valued function over $\mathcal{X}$.

**Theorem 5.** $\min_{\pi \in \mathcal{P}[\mathcal{X}]} [\mathbb{D}_\alpha(\pi_0 \parallel \pi) + \Psi_\pi^\alpha[f]] = \Psi_{\pi_0}^{\alpha-1}[f]$ with the min achieved at $\pi_{\text{co}}^*(x'|x) = \pi_0(x'|x) \exp\left(-f(x') - \Psi_{\pi_0(\cdot|x)}[-f]\right)$.

*Proof.* Suppose $\alpha \neq 0, 1$. Notice that the objective is invariant to scaling $\pi$ by any $c > 0$. Thus, we can choose a scaling to set $\sum_{x \in \mathcal{X}} \pi(x) \exp(\alpha f(x)) = 1$ so that $\Psi_\pi^\alpha[f] = 0$. Thus, the objective reduces to $\mathbb{D}_\alpha(\pi \parallel \pi_0) = \frac{\log\left(\sum_{x \in \mathcal{X}} \pi_0(x)^\alpha \pi(x)^{1-\alpha}\right)}{\alpha(\alpha-1)}$. Now, since log is monotonically increasing, we can get an equivalent problem:

$\min_\pi \ \alpha(\alpha-1) \sum_{x \in \mathcal{X}} \pi_0(x)^\alpha \pi(x)^{1-\alpha}$
Subject to $\sum_{x \in \mathcal{X}} \pi(x) \exp(\alpha f(x)) = 1, \pi(x) \geq 0$

$f(x) = a(a-1)x^{1-a}$ is always convex (it is easy to check that $f'' > 0$). Hence, the objective is convex. Since the constraint is linear, we have a convex problem. Setting the gradient of the lagrangian to 0 gives $\left(\frac{\pi_0(x)}{\pi(x)}\right)^\alpha \propto \exp(f(x))^\alpha$ so that $\pi^*(x) \propto \pi_0(x) \exp(-f(x))$. Normalizing $\pi$ to sum to 1 gives $\pi^*_{co}$ to be the optimum. Plugging in $\pi^*_{co}$ into the objective gives us $\Psi^{\alpha-1}_{\pi_0}[f]$. Taking limits as $\alpha \to 0, 1$, we get the result for all $\alpha$. □

**Lemma 1.** *Let $p, q \in \mathcal{P}[\mathcal{X}]$. Then $\mathbb{D}_\alpha(p \| q) \geq 0 \forall \alpha$, $\mathbb{D}_\alpha(p \| q) = 0 \iff p = q$. $\alpha \mathbb{D}_\alpha(p \| q)$ is a monotonically increasing function of $\alpha$*

*Proof.* $\sum_{x \in \mathcal{X}} p(x)^\alpha q(x)^{1-\alpha} = \sum_{x \in \mathcal{X}} q(x) \left(\frac{p(x)}{q(x)}\right)^\alpha$. If $\alpha \in (0, 1), x^\alpha$ in concave. Hence, by Jensen's inequality, the sum is smaller than $\leq \left(\sum_{x \in \mathcal{X}} q(x) \frac{p(x)}{q(x)}\right)^\alpha = 1$. When $\alpha \notin (0, 1)$, the reverse inequality is true since $x^\alpha$ is convex. Thus, $\log \left(\sum_{x \in \mathcal{X}} p(x)^\alpha q(x)^{1-\alpha}\right)$ is positive if $\alpha \notin (0, 1)$ and negative otherwise, showing that $\mathbb{D}_\alpha \geq 0$ for all $\alpha$. Equality holds in Jensen's if and only if every term in the sum is equal (by strict convexity/concavity) and hence $p = q$ if $\mathbb{D}_\alpha(p \| q) = 0$. Let $f(\alpha) = \alpha(\alpha - 1) \mathbb{D}_\alpha(p \| q) = \log \left(\mathrm{E}_q \left[\frac{p(x)}{q(x)}\right]^\alpha\right) = \log \left(\mathrm{E}_q \left[\exp\left(\alpha \log\left(\frac{p(x)}{q(x)}\right)\right)\right]\right)$. Letting $r(x) = \frac{p(x)^\alpha q(x)^{1-\alpha}}{\sum_{x \in \mathcal{X}} p(x)^\alpha q(x)^{1-\alpha}}$, we get

$$f'(\alpha) = \mathrm{E}_r\left[\log\left(\frac{p(x)}{q(x)}\right)\right] = -\mathrm{KL}(r \| p) + \mathrm{KL}(r \| q)$$

$$= -\mathrm{KL}(r \| p) + \mathrm{E}_r\left[\log\left(\frac{p(x)^\alpha q(x)^{-\alpha}}{\sum_{x \in \mathcal{X}} p(x)^\alpha q(x)^{1-\alpha}}\right)\right]$$

$$= -\mathrm{KL}(r \| p) + \alpha f'(\alpha) - f(\alpha) \leq \alpha f'(\alpha) - f(\alpha)$$

Thus, $(1-\alpha)f'(\alpha) + f(\alpha) \leq 0$ implying $\frac{f'(\alpha)}{1-\alpha} + \frac{f(\alpha)}{(1-\alpha)^2} \leq 0$. The LHS is just the derivative of $\frac{f(\alpha)}{1-\alpha} = -\alpha \mathbb{D}_\alpha$. Thus $\alpha \mathbb{D}_\alpha$ has a positive derivative and is monotonically increasing. □

**Lemma 2.** $\mathbb{D}_\alpha\left(\mathcal{N}(\mu_1, \Sigma^{-1}) \| \mathcal{N}(\mu_2, \Sigma^{-1})\right) = \frac{(\mu_1 - \mu_2)^T \Sigma^{-1}(\mu_1 - \mu_2)}{2}$.

Follows from simple algebra.